# Design and Performance of the GAMMA-400 Gamma-Ray Telescope for the Dark Matter Searches

A.M. Galper[1,2], O. Adriani[3], R.L. Aptekar[4], I.V. Arkhangelskaja[2],
A.I. Arkhangelskiy[2], M. Boezio[5], V. Bonvicini[5], K.A. Boyarchuk[6],
M.I. Fradkin[1], Yu.V. Gusakov[1], V.A. Kaplin[2], V.A. Kachanov[7],
M.D. Kheymits[2], A.A. Leonov[2], F. Longo[5], E.P. Mazets[4], P. Maestro[8],
P. Marrocchesi[8], I.A. Mereminskiy[2], V.V. Mikhailov[2], A.A. Moiseev[9],
E. Mocchiutti[5], N. Mori[3], I.V. Moskalenko[10], P.Yu. Naumov[2], P. Papini[3],
P. Picozza[11], V.G. Rodin[12], M.F. Runtso[2], R. Sparvoli[11], P. Spillantini[3],
S.I. Suchkov[1], M. Tavani[13], N.P. Topchiev[1*], A. Vacchi[5], E. Vannuccini[3],
Yu.T. Yurkin[2], N. Zampa[5], V.G. Zverev[2], V.N. Zirakashvili[14]

[1] Lebedev Physical Institute, Russian Academy of Sciences, Moscow, Russia
[2] National Research Nuclear University MEPhI, Moscow, Russia
[3] Istituto Nazionale di Fisica Nucleare, Sezione di Firenze and Physics Department of University of Florence, Firenze, Italy
[4] Ioffe Physical Technical Institute, Russian Academy of Sciences, St. Petersburg, Russia
[5] Istituto Nazionale di Fisica Nucleare, Sezione di Trieste, Trieste, Italy
[6] Research Institute for Electromechanics, Istra, Moscow region, Russia
[7] Institute for High Energy Physics, Protvino, Moscow region, Russia
[8] Istituto Nazionale di Fisica Nucleare, Sezione di Pisa and Physics Department of University os Siena, Siena, Italy
[9] NASA Goddard Space Flight Center and CRESST/University of Maryland, Greenbelt, Maryland, USA
[10] Hansen Experimental Physics Laboratory and Kavli Institute for Particle Astrophysics and Cosmology, Stanford University, Stanford, USA
[11] Istituto Nazionale di Fisica Nucleare, Sezione di Roma 2 and Physics Department of University of Rome "Tor Vergata", Rome, Italy
[12] Space Research Institute, Russian Academy of Sciences, Moscow, Russia
[13] Istituto Nazionale di Astrofisica – IASF and Physics Department of University of Rome "Tor Vergata", Rome, Italy
[14] Pushkov Institute of Terrestrial Magnetism, Ionosphere, and Radiowave Propagation, Troitsk, Moscow region, Russia

---

[*] Corresponding author, e-mail: tnp51@yandex.ru

**Abstract**

The GAMMA-400 gamma-ray telescope is designed to measure the fluxes of gamma rays and cosmic-ray electrons + positrons, which can be produced by annihilation or decay of the dark matter particles, as well as to survey the celestial sphere in order to study point and extended sources of gamma rays, measure energy spectra of Galactic and extragalactic diffuse gamma-ray emission, gamma-ray bursts, and gamma-ray emission from the Sun. The GAMMA-400 covers the energy range from 100 MeV to 3000 GeV. Its angular resolution is ~0.01° ($E_\gamma > 100$ GeV), the energy resolution ~1% ($E_\gamma > 10$ GeV), and the proton rejection factor ~$10^6$. GAMMA-400 will be installed on the Russian space platform Navigator. The beginning of observations is planned for 2018.

**INTRODUCTION**

In the list of very important issues in modern cosmology at the beginning of XXI century, Nobel laureate Academician V.L. Ginzburg noted "the issue of dark matter and its detection" [1]. Today is believed that the dark matter density (~25%) in the universe is several times greater than the baryonic matter density (~5%). One of the candidates for the dark matter particles are WIMPs, Weakly Interacting Massive Particles. Scientists from around the world are trying to find WIMPs, using both direct and indirect methods of detection. Indirect methods are based on the detection in cosmic-ray radiation of the annihilation or decay products of WIMPs, which can be regular particles and their anti-particles (neutrinos, electrons, positrons), as well as gamma rays. Gamma rays play an important role, as they propagate from the source without a significant absorption and, therefore, can be used to determine the direction to the source of the emission.

An analysis of observations of the region near the Galactic center by the Fermi Large Area gamma-ray Telescope (Fermi-LAT) hints for a feature in the spectrum of gamma-ray emission near ~130 GeV [2-4]. Such a feature, if confirmed, would be a unique signature of the new physics, but its reliable detection requires significant improvements in the angular and energy resolutions of the future instruments [5-9].

These challenges are addressed by a new proposed gamma-ray telescope GAMMA-400. GAMMA-400 will have a unique capability to resolve gamma-ray lines predicted to be unique signatures of the decay of WIMPs, and to determine the location of their source(s).

**GAMMA-400 GAMMA-RAY TELESCOPE**

The GAMMA-400 gamma-ray telescope is designed to measure the gamma-ray and cosmic-ray electron + positron fluxes, which may be associated with annihilation or decay of dark matter particles, as well as to survey the sky in order to search for and study gamma-ray sources, to measure the energy spectra of Galactic and extragalactic diffuse gamma-ray emission, to study gamma-ray bursts, and gamma-ray emission from the Sun in the energy range from 100 MeV to 3000 GeV.

The GAMMA-400 basic parameters are presented in Table 1. Previously described in [10, 11], the GAMMA-400 physical scheme was recently modified and is presented in Fig. 1. The GAMMA-400 gamma-ray telescope includes:

- top ($AC_{top}$) and lateral ($AC_{lat}$) anticoincidence detectors;
- converter-tracker (C), which represents 10 interleaved by tungsten (x, y) planes of silicon strip coordinate detectors with 0.1-mm pitch. The total thickness of the converter-tracker is 1.0 radiation length ($X_0$). Currently, the Italian and American scientists consider the possibility of installing to converter-tracker 15 additional (x, y) silicon planes to improve the GAMMA-400 characteristics at energies below 300 MeV;
- time-of-flight system (TOF) of S1 and S2 scintillation detectors separated by a distance of 500 mm;
- position-sensitive calorimeter, consisting of 2 parts:

(a) 4-layer imaging CC1. Each layer contains CsI(Tl) crystals and (x, y) planes of silicon strip coordinate detectors with 0.5-mm pitch. CC1 thickness is 3 $X_0$.

(b) electromagnetic CC2 from BGO crystals (25x25x250 mm$^3$). CC2 thickness is 22 $X_0$.

The total calorimeter (CC1 + CC2) thickness for the normal incidence particles is 25 $X_0$ or ~1.2 $\lambda_n$ ($\lambda_n$ is nuclear length). The total calorimeter thickness for the lateral incidence particles is ~70 $X_0$ or ~3.5 $\lambda_n$.

- S3 and S4 scintillation detectors;

- lateral calorimeter detectors (LD);

- neutron detector (ND).

Gamma-ray photons are converted into electron-positron pair in the converter-tracker, which then is detected in the instrument detectors. Anticoincidence detectors are used to identify the gamma rays, and the time-of-flight system determines the direction of the incident particles and forms the telescope aperture. Electromagnetic shower created by the electron-positron pair is developed in two parts of the calorimeter and is detected in the calorimeter and scintillation detectors S3 and S4.

Gamma rays are detected at the absence of a signal in AC, and electrons (positrons) are detected at the presence of a signal in AC, when moving downward and from lateral directions.

Using the calorimeter with thickness ~25 $X_0$ extends the particle measurable energy range up to several TeV and increases the gamma-ray telescope energy resolution up to ~1% at energies more than 10 GeV. The energy dependence of the GAMMA-400 energy resolution for incident gamma rays was simulated by the Monte Carlo method and is shown in Fig. 2 along with the same dependence for the Fermi-LAT [12] for a comparison. It is seen that in the energy range from 10 GeV to ~10 TeV the energy resolution is ~1%, which is extremely important for resolving the gamma-ray lines from the decay of the dark matter particles.

High angular resolution is achieved by determining the conversion point in the multilayer converter-tracker and the reconstruction of the shower axis in CC1. This method allows the high angular resolution of ~0.01° to be achieved at energies more than 100 GeV (Fig. 3) and enables an accurate localization of the source of the gamma-ray lines.

High-energy incident particles create a backsplash (upward moving products of the shower) in the calorimeter. To exclude the detection of the backsplash particles in AC and so manner creating self-veto, we use the method of separation of incident and backsplash particles in AC by the time of flight along with the segmentation of AC as used in Fermi-LAT [12, 13] and AGILE [14]).

The proton rejection factor of $\sim 10^6$, critical parameter for the background rejection, will be achieved by using the calorimeter and the neutron detector with other instrument subsystems.

Table 2 shows the basic parameters of the existing and planned space-based (Fermi [12], AMS-2 [15, 16]) and ground-based (MAGIC [17], H.E.S.S.-II [18], CTA [19]) experiments. It can be seen that the GAMMA-400 is well-suited for the search for the dark matter signatures including narrow gamma-ray lines.

The GAMMA-400 space observatory will be installed on the Navigator service platform designed by Lavochkin Association. It will be launched into a high-elliptic orbit with initial parameters: an apogee of 300,000 km, a perigee of 500 km, and an inclination of 51.8°. After approximately half a year the orbit will evolve in almost circular with radius of ~150,000 km, i.e., the observatory will fully leave the Earth's radiation belt. The expected lifetime of the observatory will be more than 7 years. The launch of the space observatory is planned for 2018.

**ACKNOWLEDGMENTS**

This work was supported by the Space Council of the Russian Academy of Sciences and the Russian Space Agency.

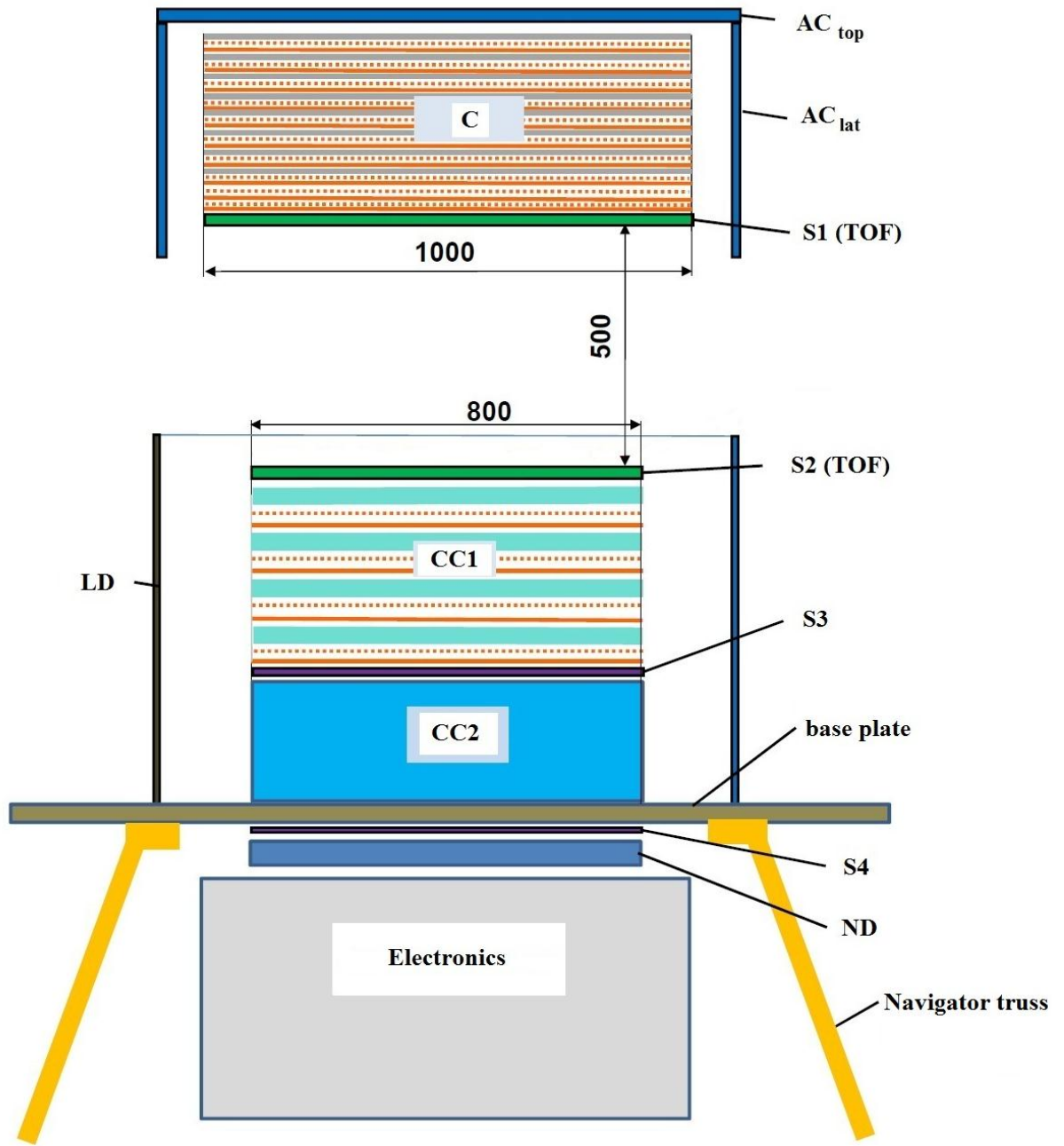

Fig. 1. GAMMA-400 physical scheme.

$AC_{top}$ is top anticoincidence scintillation detector; $AC_{lat}$ are lateral anticoincidence scintillation detectors; C is converter-tracker; S1 (TOF) and S2 (TOF) are scintillation detectors of the time-of-flight system; CC1 is imaging calorimeter; CC2 is electromagnetic calorimeter; S3 and S4 are scintillation detectors; ND is neutron detector; LD are lateral detectors.

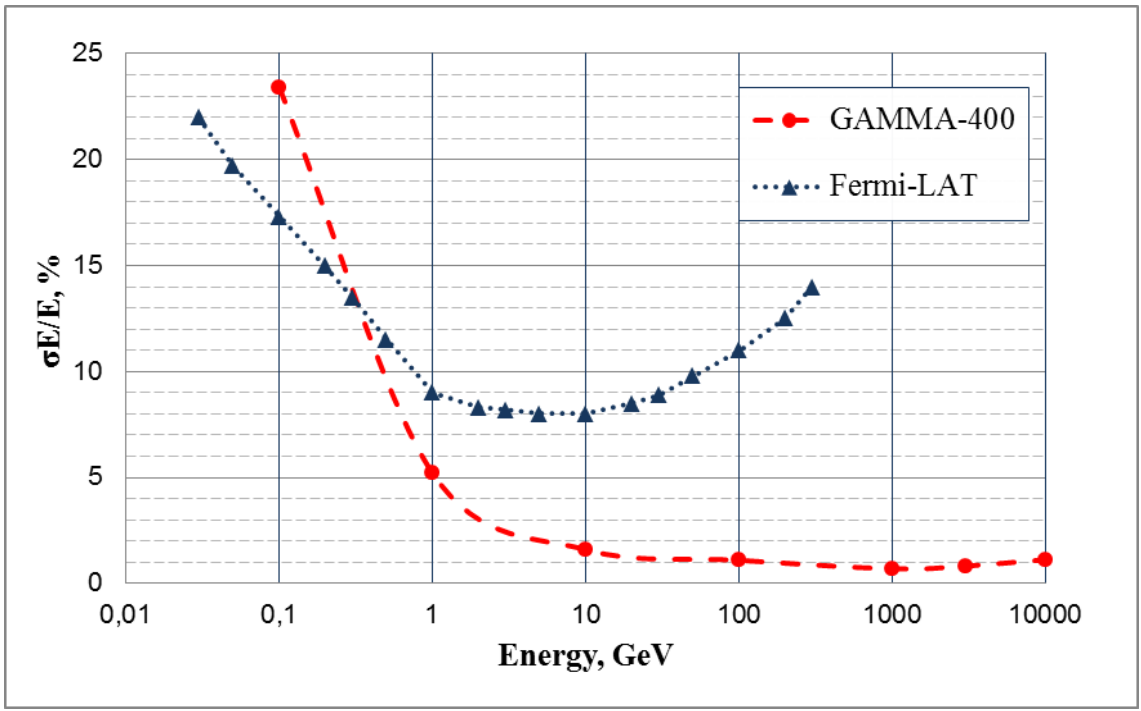

Fig. 2. Energy dependence of energy resolution for the GAMMA-400 and Fermi-LAT gamma-ray telescopes.

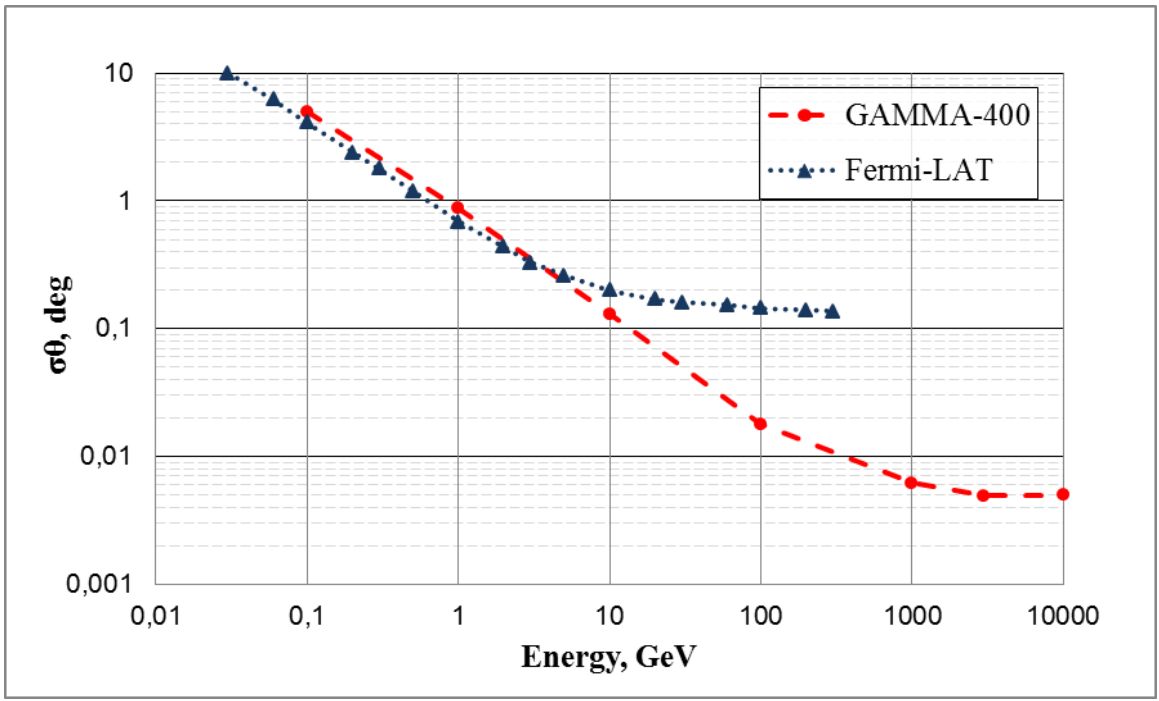

Fig. 3. Energy dependence of angular resolution for the GAMMA-400 and Fermi-LAT gamma-ray telescopes.

Table 1. GAMMA-400 basic parameters.

| | |
|---|---|
| Energy range | 100 MeV – 3,000 GeV |
| Field-of-view, sr ($E_\gamma > 1$ GeV) | ~1.2 |
| Effective area, cm$^2$ ($E_\gamma > 1$ GeV) | ~4,000 |
| Energy resolution ($E_\gamma > 10$ GeV) | ~1% |
| Angular resolution ($E_\gamma > 100$ GeV) | ~0.01º |
| Converter-tracker thickness | ~1 $X_0$ |
| Calorimeter thickness | ~25 $X_0$ |
| Proton rejection factor | ~$10^6$ |
| Telemetry downlink volume, Gbyte/day | 100 |
| Total mass, kg | 2,600 |
| Maximum dimensions, m$^3$ | 2.0x2.0x3.0 |
| Power consumption, W | 2,000 |

Table 2. A comparison of basic parameters of existing and planned space- and ground-based experiments.

|  | Space-based experiments | | | Ground-based experiments | | |
| --- | --- | --- | --- | --- | --- | --- |
|  | **Fermi** | **AMS-2** | **GAMMA-400** | **H.E.S.S.-II** | **MAGIC** | **CTA** |
| Energy range, GeV | 0.02-300 | 10-1000 | **0.1-3000** | > 30 | > 50 | > 20 |
| Field-of-view, sr | 2.4 | 0.4 | **~1.2** | 0.01 | 0.01 | 0.1 |
| Effective area, $m^2$ | 0.8 | 0.2 | **~0.4** | $10^5$ | $10^5$ | $10^6$ |
| Angular resolution ($E_\gamma > 100$ GeV) | 0.2º | 1.0º | **~0.01º** | 0.07º | 0.05º | 0.06º |
| Energy resolution ($E_\gamma > 100$ GeV) | 10% | 2% | **~1%** | 15% | 15% | 10% |